\documentclass[aip]{revtex4-1}
\usepackage{amssymb,graphicx,setspace,xspace,amsmath,appendix,natbib,tabularx,dcolumn,bm}
\usepackage{color}
\definecolor{light}{gray}{0.50}
\definecolor{heavy}{gray}{0.35}
\definecolor{black}{gray}{0.0}
\definecolor{dgreen}{rgb}{0.0,0.7,0}
\definecolor{dred}{rgb}{0.9959,0,0}
\definecolor{green}{rgb}{0.0,0.99599,0.0}
\definecolor{purple}{rgb}{0.6,0.0,0.4}

\newcounter{rupertcommentno}
\newcounter{luigicommentno}
\begin{document}
\title{Probing Spatial Locality in Ionic Liquids with the Grand Canonical
  Adaptive Resolution Molecular Dynamics Technique}
\author{B. Shadrack Jabes}
\author{C. Krekeler}
\author{R. Klein}
\author{L. Delle Site}
\date{\it{\today}}
\email[Author for correspondence:]{luigi.dellesite@fu-berlin-de (L.D.S)}
\affiliation{Institute for Mathematics, Freie Universitat Berlin, D-14195 Berlin, Germany}
\begin{abstract}
We employ the Grand Canonical Adaptive Resolution Molecular Dynamics Technique (GC-AdResS) to test the spatial locality of the 1-ethyl 3-methyl imidazolium chloride liquid. In GC-AdResS atomistic details are kept only in an open sub-region of the system while the environment is treated at coarse-grained level, thus if spatial quantities calculated in such a sub-region agree with the equivalent quantities calculated in a full atomistic simulation then the atomistic degrees of freedom outside the sub-region play a negligible role. The size of the sub-region fixes the degree of spatial locality of a certain quantity. We show that even for sub-regions whose radius corresponds to the size of a few molecules, spatial properties are reasonably {reproduced} thus suggesting a higher degree of spatial locality, a hypothesis put forward also by other {researchers} and that seems to play an important role for the characterization of fundamental properties of a large class of ionic liquids.
\end{abstract}
\maketitle
\section{Introduction}
Ionic liquids (ILs) are a new class of air- and water-stable organic
salts whose melting point is around room temperature\cite{sw16}. They have the ionic nature
of inorganic solvents and organic nature of organic solvents. Because of this
dual or amphiphilic nature of ions, they inherit both the solvent characteristics and possess even more excellent properties, such as high electrical {conductivity}, improved electrical
and thermal {stability}, good solvation ability, negligible vapor {pressure},
non-flammability, to mention a few\cite{ps08,hl08}. 
When the balance between the dual nature of ILs is altered by chemical
modifications of the polar or apolar components or charge of the ions, an
exponentially large number of ILs with a wide range of physicochemical properties  such as viscosity,
catalytic activity, or solvation can be generated. 
Ionic liquids engineered at the molecular level to have specific physical
 and chemical properties, can be employed for a range of designated
 applications\cite{rg10,srds11}: ionic liquids as
drugs\cite{swnsfcsp15,hdar13}, as batteries and
super-capacitors for solving energy
demands of the environment\cite{lglw15,ghznwz15,pelw15,zymp14}, to capture greenhouse gases, to dissolve
proteins\cite{tas15,gsk10} and cellulose\cite{spsp14,wgr12}. 

{The design and testing of new liquids, however,} given the innumerable chemical
combinations of anions and cations, rises the problem of {their} optimal design
since using experiments  is a  rather time consuming and
unfeasible strategy \cite{mccgv11}. Therefore, many studies involve
the use of computer simulations so as to understand
and predict such properties {\it in silico} and then drive experiments{, and the
present paper contributes to the development of suitable simulation technology}.

The key question that chemical engineers must face for ILs concerns the
possibility of linking, in a rational way, local
structural organization or microscopic interactions and the large scale physico-chemical behavior. 
For example: the separation of fission products such as Cs(+) and Sr(2+) from aqueous solutions 
using chemically modified ionic liquids are influenced by the structures of imidazolium cations 
attached to each other by covalent bonding\cite{ldbb06}. The hydroxyl- and carboxyl-functionalised 
ILs {exhibit} molecular recognition characteristics that can be used to separate even 
homologue zwitterionic phospholipid species out of homogeneous mixtures. {These} selectivity
characteristics can be augmented by the tendency of ILs to form meso-phases, polar-apolar 
alternations at the nanoscale\cite{ur04,wv05,cp06,wv06,trbd07}. Thus a profound understanding 
{of local structures of ILs in the} bulk or at interfaces and the relationship between
local structures and thermodynamic properties is mandatory for choosing an optimal combination 
of ILs {to obtain the desired} properties on demand.
 
Several computational studies have been carried out to characterize the locality in ILs, that 
is, how a molecule and its immediate {surroundings} respond to changes occurring 
{on} the global mesoscopic scale, e.g., the hydrogen bonding network of the bulk or other 
properties which typically occur on a larger scale \cite{skdzbdh10,kdszhbd10,wv06,wv05,wzdbhkd11}. 
In particular, the structural locality and its relation to the governing interactions  was 
actually either studied at a quantum level or at a coarse grained level. For example: {Ab 
initio simulation} studies suggest that the electric dipole moment of the ions is highly local, 
meaning that it is sufficient to know only the details of the very immediate environment 
\cite{wzdbhkd11}. {At the same time,} coarse-grained simulations have been used to study the 
heterogeneous structural ordering and its dynamics providing a very simplified model of the 
bulk \cite{wv06,wv05}.

For the studies reported above, {however,} one must consider that {\it ab initio} 
techniques are limited to small gas phase systems of relatively small samples of liquids, 
{and are thus} not yet appropriate for the exhaustive sampling of large scale liquid 
properties. Coarse-grained models{, in turn,} use {a} simplified approach 
{that ignores the role of} atomistic degrees of freedom and thus {allows} the 
chemical nature of the anions and cations {to enter} only marginally. A reasonable 
compromise for linking the atomistic and mesoscopic scale is the use of classical Molecular 
Dynamics (MD) simulations {in which} the only limiting requirement is the necessity 
of long simulations times for a proper statistical sampling. In {present} work, in 
particular, we employ classical molecular dynamics within the context of the 
Adaptive Resolution Simulation approach (AdResS) \cite{adress2005,annurev2008} in its 
Grand Canonical version (GC-AdResS) \cite{wss12,whss13,azhws15}.

GC-AdResS is a multi-scale method that allows the molecules to change their resolution on 
the fly from high-resolution to lower resolution region and vice versa.  The higher 
resolution region corresponds to a local region in space that is defined by an {atomistically} resolved model while the surrounding 
bulk requires a less accurate model (coarse-grained), sufficient for reproducing only the 
large scale thermodynamic properties. In essence, the region of interest {inside of which} 
atomistic details are relevant is embedded in a thermodynamic bath with which {it} 
exchanges energy and particles in a Grand Canonical fashion. Relevant technical details of 
GC-AdResS are reported in the technical section; for a general overview of this method and 
related ones see Ref.\cite{physrep2017}. 

The {key aspect of GC-AdResS that is important in the context of the present work} is
its capability to be used as an analysis tool of MD simulations to identify, for a given 
property, the relevant atomistic degrees of freedom of the system for a certain quantity or 
property of interest. In fact, a systematic variation of the size of the {atomistically} 
resolved region implies a systematic variation of the number of atomistic degrees of freedom. Next, 
suppose we compare a given property calculated in the atomistic region of GC-AdResS with the same 
property calculated in an equivalent sub-region of a full atomistic simulation. {By such a
comparison it is possible to draw unambiguous conclusions regarding the influence of the {explicit} atomistic 
degrees of freedom residing outside of the sub-region on our property 
of interest}. In fact if the property calculated shows the same behavior in both cases, it implies 
that the role of the {explicit} atomistic degrees of freedom outside the {considered sub-region} is 
negligible. This conclusion in turn implies that the minimum size of a sub-region where
GC-AdResS and a full atomistic simulation agree (for a certain quantity)
defines the degree of spatial locality of such quantity w.r.t. the atomistic
degrees of freedom. 

AdResS as a tool to identify spatial locality in complex molecular systems has been already 
successfully applied in past and recent work (see Refs.\cite{jcp-2010-full,pccp-2017-full}). 
In the context of ILs, coming back to the earlier discussion about the necessity of linking 
specific chemical structures to large scale properties for an optimal design and combination 
of anions and cations, the GC-AdResS analysis can play a relevant role. In fact, its aim is 
exactly the identification of the minimal high resolution islands of anions and cations
which, without loss of accuracy, can be embedded in the larger bulk with its mesoscopic 
properties obtained at lower resolution. In this work we {provide} a practical example 
for a relevant IL, [Emim][Cl]. Of particular interest is the result that the degree of
locality can be much higher than one may expect at {an} intuitive level. The 
satisfactory technical applicability of GC-AdResS to ILs, employing charged or uncharged 
spherical particles as coarse-grained models, has already been proven, in a rather detailed 
manner, in preceding work \cite{ks17}. {Therefore,} this paper represents
{an} extension of the technical implementation to the use as a multi-scale analysis 
tool. A discussion {regarding} the utility of information {on} locality for 
chemical design and molecular modeling in ILs {concludes} the paper.

\section{Computational details}
\subsection{Method}
\begin{figure}[htbp]
	        \centering
                \includegraphics[clip=true,trim=0.1cm 0cm 0cm
                0.1cm,width=10cm]{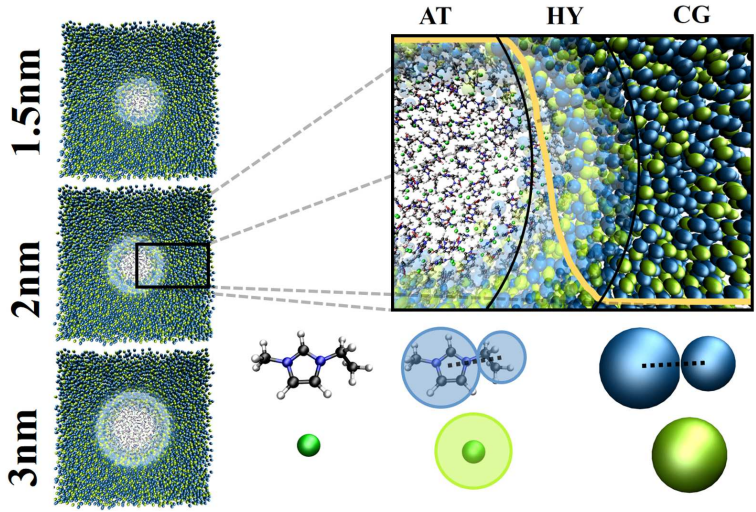}
                \caption{(Right column)Schematic representation of the GC-AdResS scheme;
                CG, HY and AT represents the coarse-grained region, hybrid
                region and the atomistic region respectively. {The CG model consists of one sphere for the imidazolium ring and one sphere for the alkyl side chain (cation) and a sphere for the anion which has an effective size different from the atomistic representation.} Within the AT and
                CG regions the force acting between two molecules is derived
                from the respective potentials. However, in the HY region, the
                forces are obtained by a space dependent interpolation scheme in
                which a weighting function w(x) (shown in yellow) allows a
              smooth transition from one resolution to another. In the left column we show three
            AdResS simulations used in the study with three atomistic resolutions {spanning from 1.5nm to
          3nm for the radius of the atomistic region}}
                \label{fig:fig1}
                \end{figure}

We employ the Adaptive Resolution Simulation Scheme (AdResS)
\cite{adress2005,annurev2008} for molecular dynamics in its more advanced
version in the Grand Canonical interpretation (GC-AdResS)
\cite{wss12,whss13}.
AdResS is a multi-resolution simulation method that links two regions of a
simulation box having
different resolution, regions:1. atomistic (AT) and 2. a coarse-grained (CG)
system, and it allows the free exchange of particles between the two regions
through a coupling transition region (HY) where molecules have space-dependent
hybrid atomistic/coarse-grained resolution.
In the hybrid region, the force between two molecules $\alpha$ and $\beta$ is computed via a space-dependent
interpolation formula and is written as
\begin{equation}
  F_{\alpha \beta} = w(X_{\alpha})w(X_{\beta})F_{\alpha \beta}^{AT} + [1-
  w(X_{\alpha})w(X_{\beta})]F_{\alpha \beta}^{CG}
\end{equation}
here, $F_{\alpha \beta}^{AT}$ is the force between the particles derived using
the atomistic potential and  $F_{\alpha \beta}^{CG}$ is the force derived from
the CG potential. $w(x)$ is the interpolating switch and is defined as:
\[ 
  \left \{
      \begin{tabular}{ll}
        1 & $x < d_{AT}$ \\
        $cos^2[{\pi}{2(d_\Delta)}(x-d_{AT})]$ & $d_{AT} < x < d_{AT} +
        d_{\Delta}$ \\
        0 & $d_{AT}+d_{\Delta}< x$ 
          \end{tabular}
        \right \}
\]
where, $d_{AT}$ and $d_{\Delta}$ are the spatial extent of the atomistic region
and hybrid regions respectively (see Fig.\ref{fig:fig1}). The
weighting function (shown in yellow line) smoothly goes from 0 to 1  in the
hybrid region and allows the coarse grained particles to change their resolution into an atomistic
molecule and vice versa. 
This minimal set up has been shown already sufficient for performing accurate
simulations in both the atomistic and coarse-grained region (see e.g.\cite{water-2008}). 

{However,} further technical and conceptual developments led to the definition of a
thermodynamic force in the transition region which assures {\it a priori}
proper thermodynamic equilibrium and correct particles' exchange between he AT
and CG region \cite{jcp2010-simon,prl2012}. Furthermore, it
has been shown that for the coarse-grained model it is sufficient to develop
models that reproduce only density and temperature, that is macroscopic
thermodynamics, so that AdResS is re-framed within the Grand Canonical
approach \cite{wss12,whss13}, or even further, within the more
general Grand Ensemble idea of open boundary systems \cite{azhws15,luigi2016,physrep2017}. 
For the specific study in this work, the atomistic region is
modelled as a sphere of radius $d_{at}$ and is surrounded by a spherical hybrid region of width 
$d_{hy}$ and the
remainder of the box is the CG region that serves as a particle reservoir for
the AT region; see the pictorial representation in Fig.\ref{fig:fig1}. The force field
parameters of the atomistic model and of the coarse-grained model are reported
in sections \ref{fig:models} and \ref{fig:tdet}. 
\subsection{Models}
\label{fig:models}
The full atomistic 
force field parameters for modeling 1-ethyl 3-methyl imidazolium chloride were taken from
a study by Dommert et al \cite{dwbsh12}, {in particular such model considers reduced/scaled charges of  +0.8e and -0.8e instead of the full formal charges +1.0 e and -1.0e}. In the coarse-grained region, ILs
are modelled as charged/neutral spherical beads, {i.e. one sphere for the imidazolium ring, bonded to one sphere for the alkyl side chain, for the cation and a sphere with an effective excluded volume, consistent with the CG model of the cation, for the anion (see Fig.\ref{fig:fig1}). The interaction parameters are developed using a straightforward Inverse Boltzmann Iterative
procedure (IBI).} Otherwise care must be taken in deriving the generic CG model
because a charge neutral representation of coarse-grained bead for ILs would
result in gaining artificial electrostatic interactions instantaneously in the atomistic and
the reservoir regions. The IBI procedure employed in this study takes into
account the corresponding charge in the
coarse-grained model during the iterative process\cite{ks17}. 
\subsection{Technical details}
\label{fig:tdet}
All  the simulations were performed using GROMACS package\cite{hkvl08}. 
We set up two systems, the first one
(400 ion pairs) was used to derive two coarse grain potentials, which we then
transferred to the larger system (19968 ion pairs). We optimized both systems
using full atomistic NPT calculations. The simulation temperature was set to
400K, the time step was 2fs and the electrostatic interactions were calculated
through the particle mesh Ewald (PME) technique. For the first 5 ns, we used
Berendsen barostat, \cite{bpv84} after that we switched to the
Parrinello-Rahman barostat\cite{pr81}. for the following 5ns. We
monitored the box size and considered that we have reached convergence, when the
changes in the box length were of the order of 0.0001nm. 
For 400 and 19968 ion pairs we obtained a
cubic box with 4.31080 nm and 15.81094 nm respectively. Furthermore, the radial
distribution function were obtained after 20ns full atomistic NVT simulations at
380K and with 2fs timesteps. 
The 400 ion pairs configuration was used to derive the
two coarse-grained models. We used an inverse Boltzmann iteration, IBI,
procedure \cite{rpm03} to reproduce the radial distribution functions of the full
atomistic target systems (for charged model see Fig.
\ref{fig:fig3}-\ref{fig:fig5}
and for the neutral model see Fig.
\ref{fig:fig6}-\ref{fig:fig8}).
After we derive these tabulated potentials, we used the configuration from the
19968 ion pairs to set up the GC-AdResS system. We vary the resolution of the
atomistic region in the GC-AdResS simulation from 1.3nm - 3nm in order to look at the
essential degrees of freedom that are responsible to capture the structural
properties of the system similar to the full atomistic simulations. This atomistic
region is bordered by hybrid regions of length 2nm, while the remaining length
corresponds to the coarse-graining region. As in standard AdResS simulations, a
Langevin thermostat is used with $\Gamma = 5 ps^{-1}$. The electrostatic
interactions, as usually done in AdResS, were treated by the generalized
reaction field method with a self-consistent dielectric constant as calculated
by GROMACS\cite{fjk12}. The use of the charged coarse-grained model
implied that the reaction field method is also used in the coarse-graining
region where it applies in the same way as in the atomistic region. The
thermodynamic force through the iterative procedure was considered converged
after 100 iterations, sufficient to reach an accuracy of $<5\%$ particle
density and for the ion-ion radial distribution functions ( compared to the full
atomistic of the reference). 
               \begin{figure}[htbp]
	        \centering
                \includegraphics[clip=true,trim=0cm 0cm 0cm
                0cm,width=8cm]{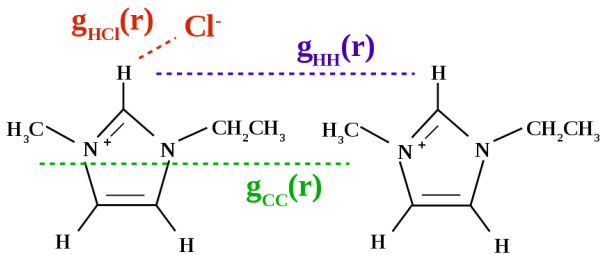}
                \caption{Pictorial representation of the atom types
                  corresponding to the radial distribution profile for the emim chloride system. The center of mass
            of the cations and anions are represented as CAT,
          ANN or Cl respectively. The carbon (C), of the methyl group is also
        represented as CAA.}
                \label{fig:fig2}
                \end{figure}

                \begin{figure}[htbp]
	        \centering
                \includegraphics[clip=true,trim=0cm 0cm 0cm
                0cm,width=8cm]{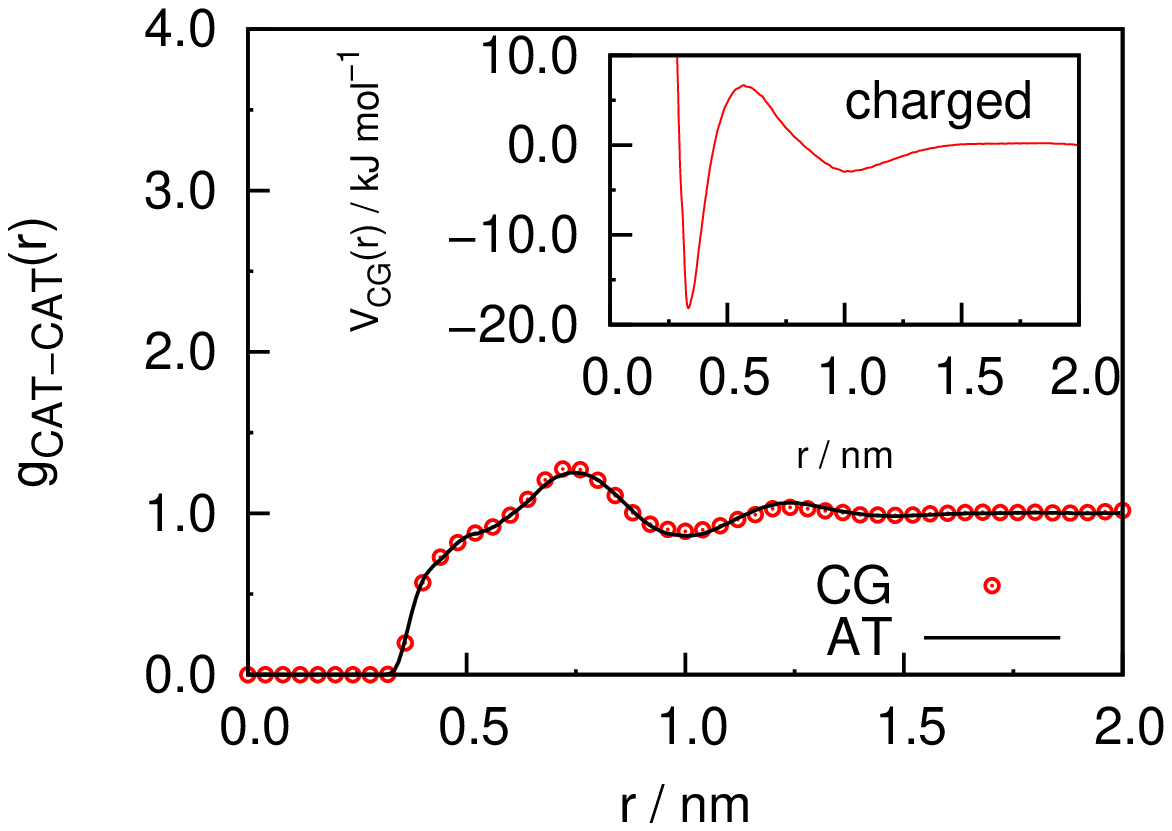}
                \caption{The cation-cation radial distribution
                profile obtained from atomistic simulation (AT) and
              coarse-grained (CG) simulation as a function of distance $r$. The
              CG simulation employs the IBI potential for the charged CG
            model (inset).}
                \label{fig:fig3}
                \end{figure}

                \begin{figure}[htbp]
	        \centering
                \includegraphics[clip=true,trim=0cm 0cm 0cm
                0cm,width=8cm]{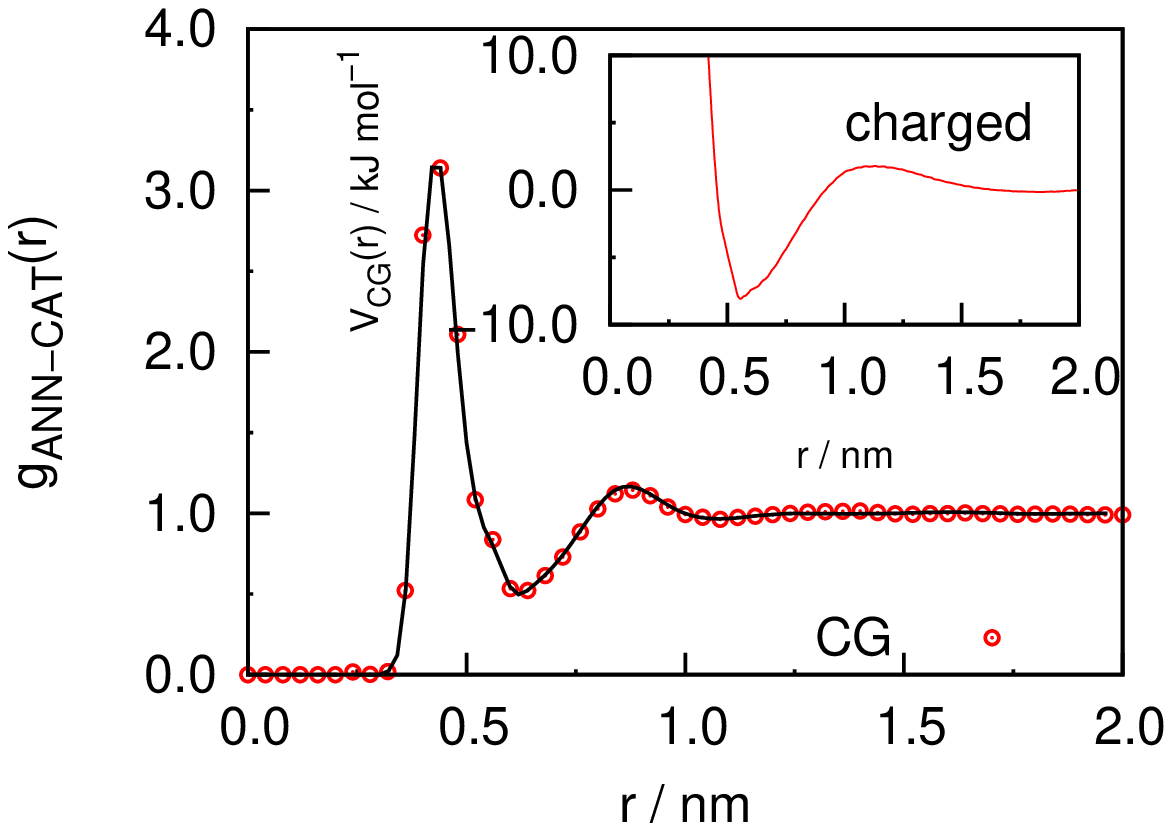}
                \caption{The anion-cation radial distribution
                profile obtained from atomistic simulation (AT) and
              coarse-grained (CG) simulation as a function of distance $r$. The
              CG simulation employs the IBI potential for the charged CG
            model (inset).}
                \label{fig:fig4}
                \end{figure}

                \begin{figure}[htbp]
	        \centering
                \includegraphics[clip=true,trim=0cm 0cm 0cm
                0cm,width=8cm]{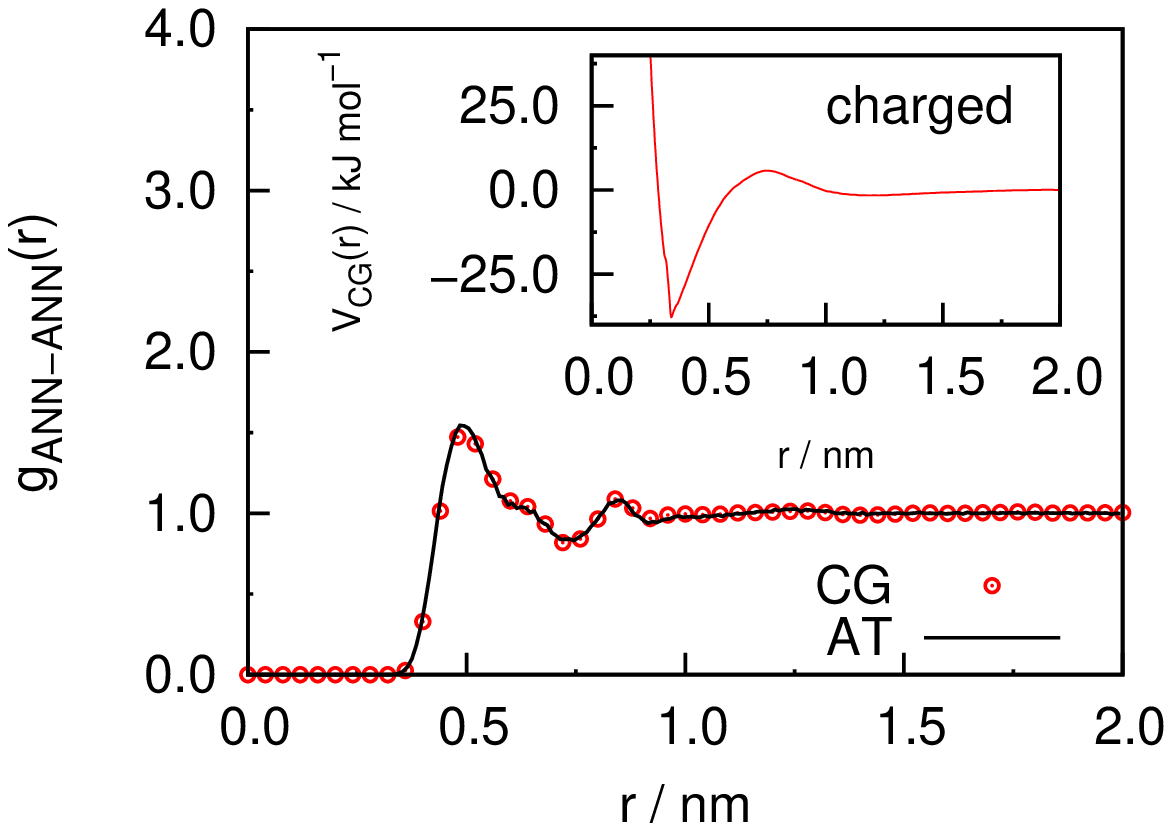}
                \caption{The anion-anion radial distribution
                profile obtained from atomistic simulation (AT) and
              coarse-grained (CG) simulation as a function of distance $r$. The
              CG simulation employs the IBI potential for the charged CG
            model (inset).}
                \label{fig:fig5}
                \end{figure}

                \begin{figure}[htbp]
	        \centering
                \includegraphics[clip=true,trim=0cm 0cm 0cm
                0cm,width=8cm]{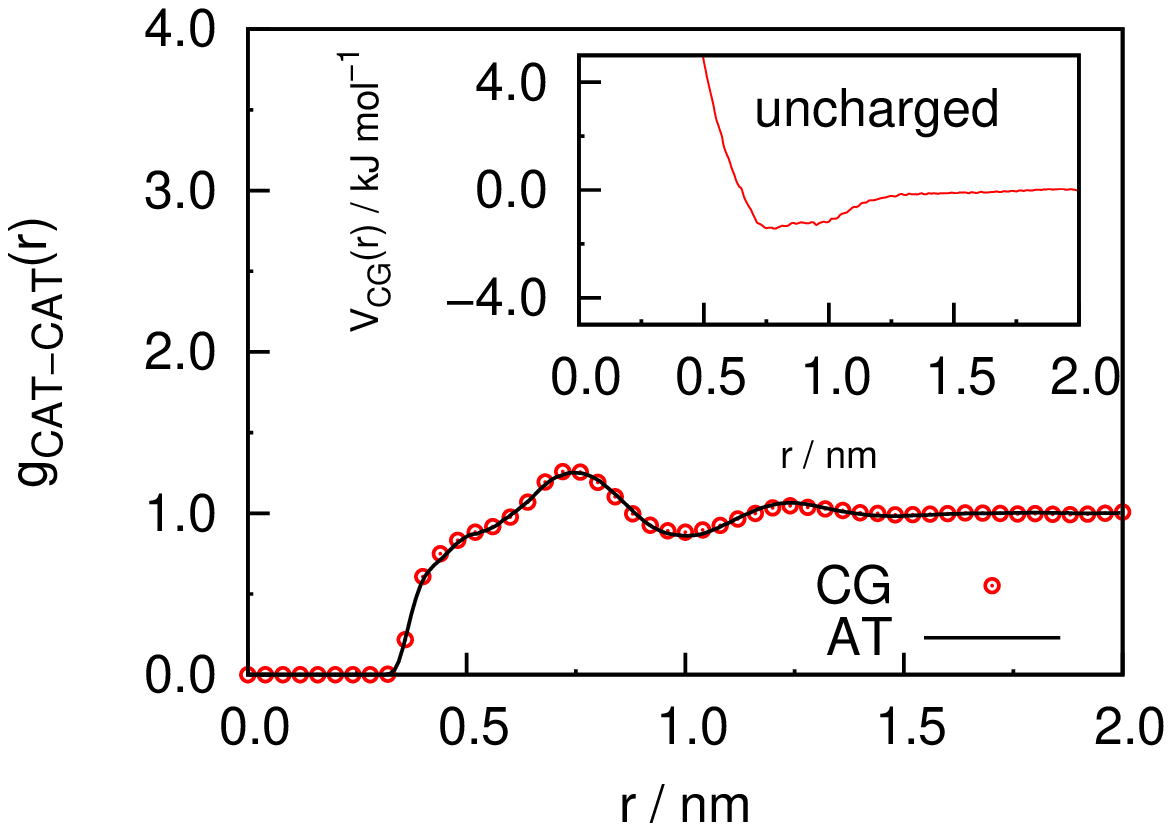}
                \caption{The cation-cation radial distribution
                profile obtained from atomistic simulation (AT) and
              coarse-grained (CG) simulation as a function of distance $r$. The
              CG simulation employs the IBI potential for the uncharged CG
            model (inset).}
                \label{fig:fig6}
                \end{figure}

                \begin{figure}[htbp]
	        \centering
                \includegraphics[clip=true,trim=0cm 0cm 0cm
                0cm,width=8cm]{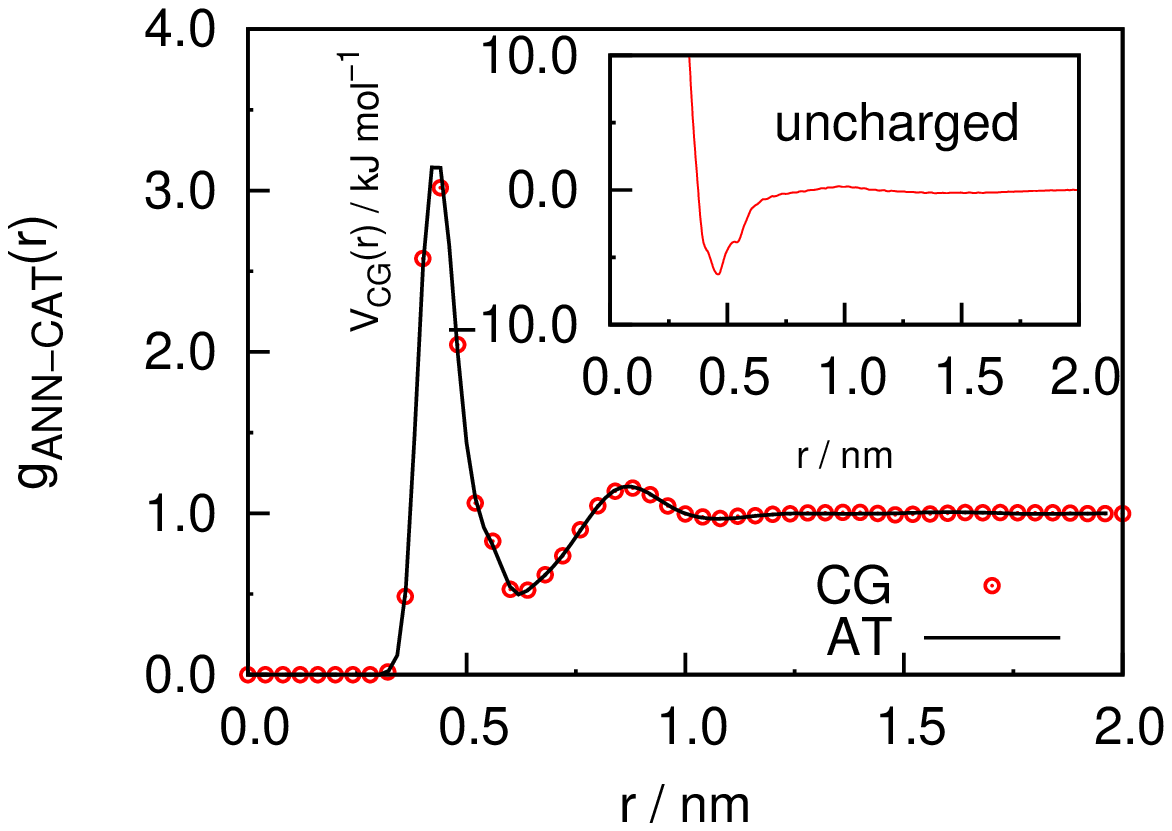}
                \caption{The anion-cation radial distribution
                profile obtained from atomistic simulation (AT) and
              coarse-grained (CG) simulation as a function of distance $r$. The
              CG simulation employs the IBI potential for the uncharged CG
            model (inset).}
                \label{fig:fig7}
                \end{figure}

                \begin{figure}[htbp]
	        \centering
                \includegraphics[clip=true,trim=0cm 0cm 0cm
                0cm,width=8cm]{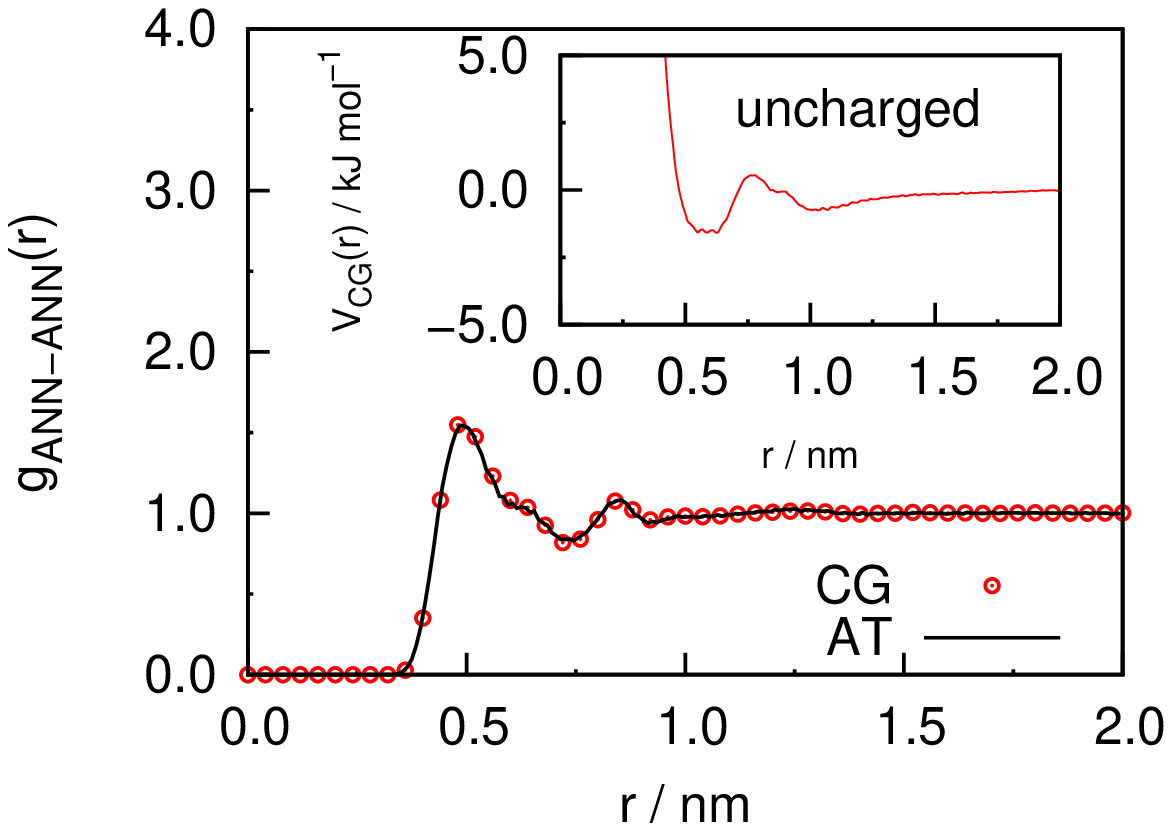}
                \caption{The anion-anion radial distribution
                profile obtained from atomistic simulation (AT) and
              coarse-grained (CG) simulation as a function of distance $r$. The
              CG simulation employs the IBI potential for the uncharged CG
            model (inset).}
                \label{fig:fig8}
                \end{figure}
\section{Results and discussion}
The GC-AdResS analysis involves two stages, the first concerns a careful check
that necessary conditions of physical consistency are fulfilled, the second
concerns the analysis of spatial locality done through quantities which
capture the essence of the statistical properties of a system. In the first stage
we check that the particle number density of the AdResS system agrees well
(within a chosen accuracy, usually with $3-5 \%$, which has been already shown
in previous work to be satisfactory); if this condition would not hold than we can
obviously not have the same thermodynamic condition of the reference full
atomistic system. Next we must check that the particle number probability
distribution function of the atomistic region agrees well (once again within the
chosen accuracy) with the same quantity calculated in the sub-region of a full
atomistic simulation with the same size of the atomistic region considered in GC-AdResS.
Such a quantity expresses the behavior of the particle number fluctuation 
and thus assures that the exchange of matter with the large reservoir occurs
in the correct way. Finally, in order to assure that artificial structural effects due to
the hybrid region are negligible in the atomistic region, we calculate the
probability distribution of an order parameter and compare with the equivalent
calculated in the sub-region of the full atomistic region.
The order parameter is system specific, while the other two quantities are
universal; together they significantly characterize the system under investigation.

Once we are assured about the proper thermodynamic and statistical behavior
of the atomistic region, we proceed with the analysis of the locality. For
such an analysis we consider several radial distribution functions calculated
in the atomistic region and {compare them} with the equivalent {functions} 
in the corresponding sub-region of the full atomistic system. In fact radial
distribution functions represent the $3N$-probability distribution function of the
system at the second order, where the order refers to the reduction of the
$3N$-probability distribution function to a factorized function of
independent two-body terms (see also \cite{whss13}). If the radial distribution
functions of the atomistic region of GC-AdResS reproduce the behavior of the
same quantities calculated in the sub region of a full atomistic simulation,
then the two sub-regions are statistically equivalent at least up to the second
order. That is, within such an accuracy, the atomistic degrees of freedom outside 
the sub-region are not relevant for determining quantities and properties of such a 
sub-region. {In fact the atom-atom radial distribution functions of the hybrid region cannot reproduce the radial distribution function of the corresponding full atomistic simulation of reference (see \cite{whss13}). Thus if atom-atom radial distribution functions in the AT region of AdResS reproduce the results of the reference full atomistic simulation, it implies that the only relevant atomistic degrees of freedom are those of the AT region.} In practice the agreement between GC-AdResS and the full atomistic simulation
regarding the local radial distribution functions tell us  that
the ensemble average of quantities depending on spatial coordinates can be
localized in the sub-region with an accuracy at least up to the second order
in the probability distribution. It has been shown that the second order is
sufficient in MD to have a high accuracy for several relevant quantities and
properties (see Ref.\cite{whss13}).

\subsubsection{Consistency of GC-AdResS as a open boundary system}
\begin{figure}[htbp]
	        \centering
                \includegraphics[clip=true,trim=6cm 11.8cm 6cm
                3cm,width=8cm]{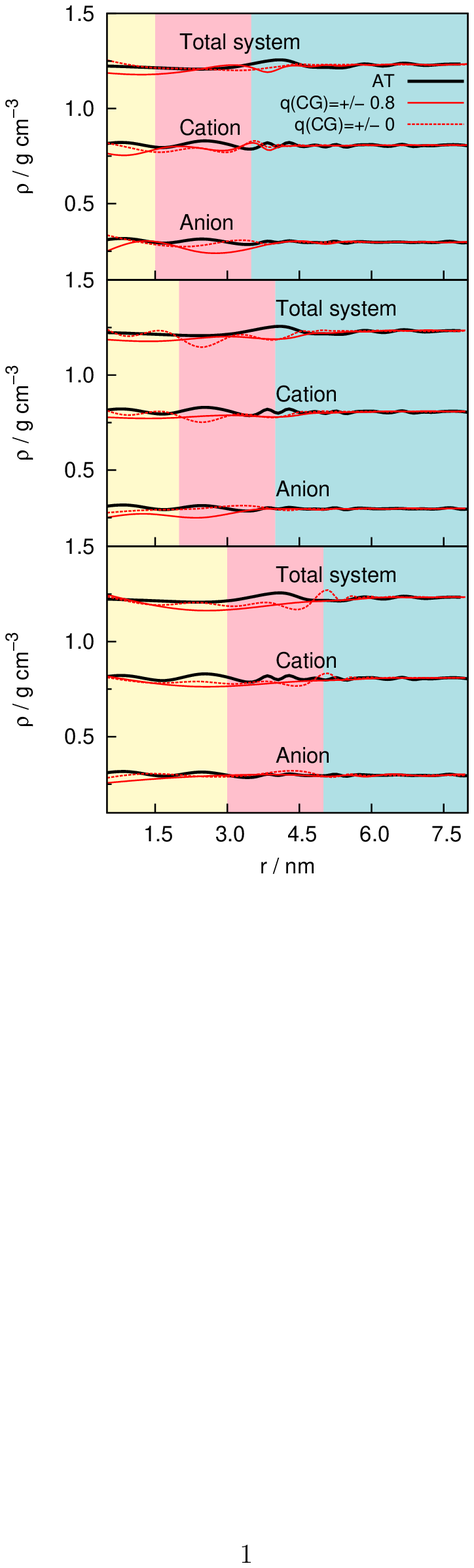}
                \caption{The particle number density of ions pairs as a function of the distance from the center of the atomistic region. The yellow,
                pink and blue colors corresponds to atomistic, hybrid and
              coarse grained regions in the AdResS simulation. The experimental
            density of [Emim][cl] is 1.2g/$cm^3$\cite{kfmm14}.}
                \label{fig:fig9}
                \end{figure}

Figure \ref{fig:fig9} shows the particle number density of ions pairs as a function of the distance 
from the center of the atomistic region. We have considered three different atomistic regions of 
radius $1.5 nm$, $2.0 nm$ and $3.0 nm$ respectively. The results reported in Fig.\ref{fig:fig9} show 
that the behavior of the ion pairs density for each atomistic region considered in GC-AdResS is 
satisfactory, independently from the coarse-grained model. Such an accuracy implies that the overall 
thermodynamic state point ($\rho$, T) of the GC-AdResS simulations is the same {as that} of 
the reference full atomistic simulation {within acceptable bounds}. 

\begin{figure}[htbp]
\centering
\includegraphics[clip=true,trim=6cm 9.8cm 6cm 3cm,width=8cm]{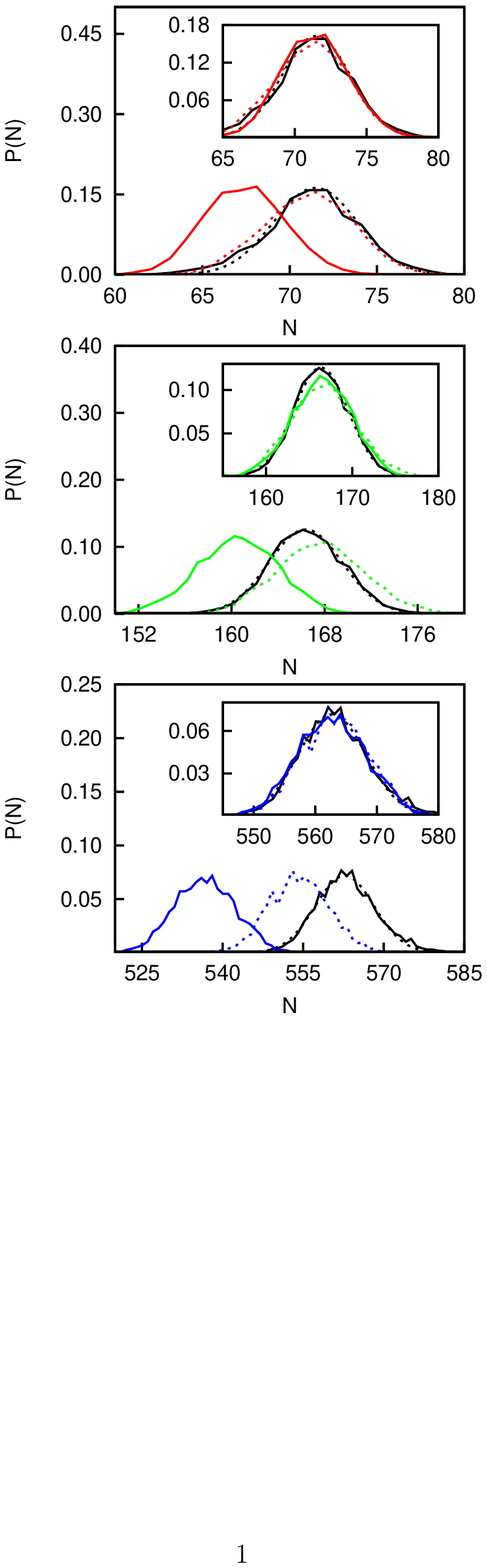}
\caption{The probability distribution of the ion pairs number density in the atomistic region 
of the three AdResS simulations (top 1.5nm, middle 2nm and bottom 3nm) compared with the 
corresponding quantity in the equivalent sub-region of the full atomistic simulation. The 
charged and neutral CG models are represented in lines and dashed lines respectively. The 
distribution profile for the reference full atomistic system is shown in black solid line. 
The Gaussian fit for the reference system is shown in black dashed line. The location of the 
maximum of the distribution obtained in GC-AdResS simulations is shifted compared to the 
full atomistic reference. In fact we have set the accuracy of our calculations,through 
the thermodynamic force, at $5\%$  and the maximum shift found is below $5\%$. In such a 
case what is more relevant is the Gaussian shape of the function and from the insets, where 
the functions are over-imposed to the Gaussian fit, one can see that the agreement is satisfactory.}
\label{fig:fig10}
\end{figure}

Figure \ref{fig:fig10} shows the probability distribution of the ion pairs
number density in the atomistic region of the three AdResS simulations compared
with the results obtained for the corresponding sub-regions of the full
atomistic simulation. The systematic shift in the location of the maxima, w.r.t.
the full atomistic case, is due to the accuracy chosen for the thermodynamic
force. We have chosen a threshold of $5\%$ and in fact the largest shift in
Figure \ref{fig:fig10} does not go beyond the threshold chosen. Our choice of
$5\%$ is based on the experience of previous studies, where we have seen that
such an accuracy is {sufficient} for obtained satisfactory results for structural
properties in the atomistic region (see for example the discussion about accuracy of density in Refs.\cite{jcp-2015-pi,peters}). If a higher accuracy is needed, it would be
sufficient to run the thermodynamic force with stricter criteria of convergence.
The results of Figure \ref{fig:fig10} show a proper behavior of the probability distribution, which 
implies that the statistical exchange of molecules with the reservoir consisting of the hybrid and 
coarse-grained region occurs according to the expectations of a well behaving system with open 
boundaries. As for the density, the satisfactory character of the agreement with reference data does 
not depend strongly on the coarse-grained model. In fact the results obtained with the two 
{model} show differences which are within the accuracy of the calculations.

\begin{figure}[htbp]
	        \centering
                \includegraphics[clip=true,trim=6cm 8cm 6cm
                3cm,width=8cm]{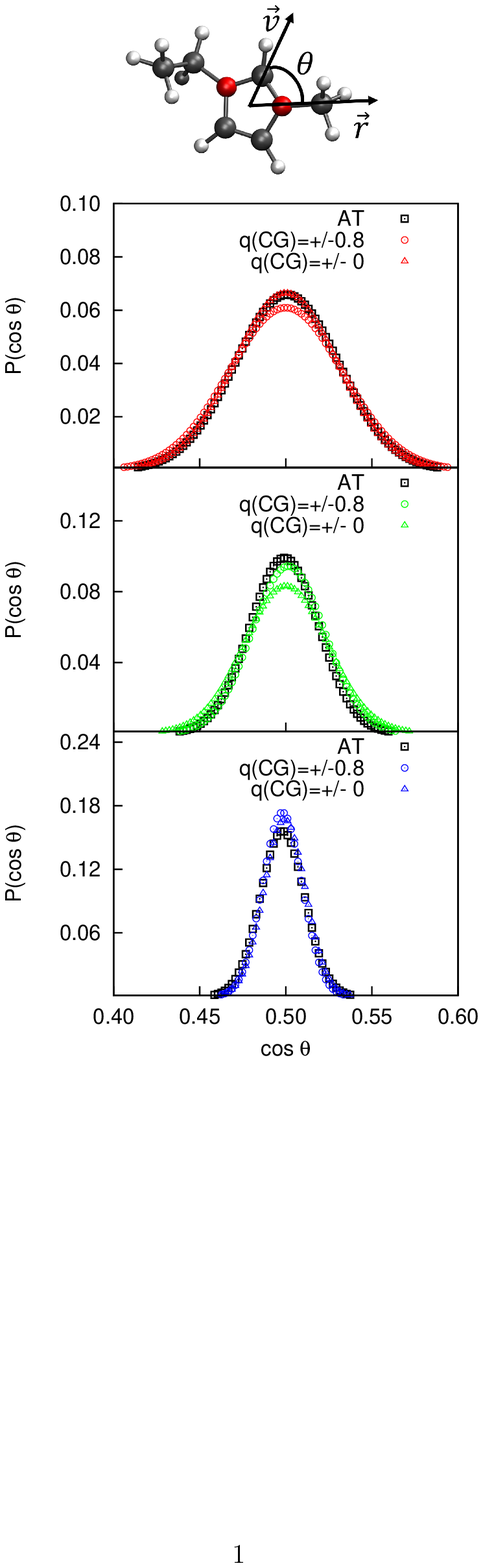}
                \caption{The probability distribution of the averaged order 
                parameter $|cos\theta|$ (top panel) in the atomistic region of the three AdResS simulations (top 1.5nm, middle 2nm and bottom 3nm) compared with the corresponding quantity in the equivalent sub-region of the full atomistic simulation}
                \label{fig:fig11}
              \end{figure}

Finally, Figure \ref{fig:fig11} shows the probability distribution of the averaged order parameter 
$|cos\theta|$, where $\theta$ is defined as the angle between the symmetry vector of the cation 
and the radial unit vector $\hat{r}$ with origin at the center of the spherical atomistic region. 
Also in this case the agreement is satisfactory. The charged coarse-grained model for 
$1.5 \, {\rm nm}$ and $2.0  \, {\rm nm}$ is slightly more accurate that the uncharged model.
{However,} the difference between the two cases is, in general, not size-able. Such 
results show that the artificial change of resolution along the radial direction does not imply an 
artificial alignment of the cations due to the hybrid interface.
The calculation of the quantities reported in this section assures as that the three subsystems we are considering in GC-AdResS properly behave at the relevant statistical and structural level and thus legitimate to go a step forward and perform the analysis of locality on more detailed and specific quantities, that is the radial distributions functions.
\subsubsection{Probing spatial locality}
As anticipated above, in order to probe the degree of spatial locality we
calculate various radial distribution functions since they represent the
$3N$-probability distribution function of the system at the second order.
Moreover, the functions considered here are not only those defined by centers of
mass of the cations but also those defined between different specific atoms.
This means that they reproduce at the second order very detailed aspects of the
$3N$-probability distribution. {Thus,} if the atomistic region of GC-AdResS
accurately reproduces such distributions then {we infer that} the
atomistic degrees of freedom outside the atomistic region do not play a relevant
role in the atomistic structure of such a region. As a consequence the
calculation of space-dependent physical quantities will need only the knowledge
of the local atomistic environment, \emph{i.e.}, the atomistic region. 

\begin{figure}[htbp]
	        \centering
                \includegraphics[clip=true,trim=0cm 0cm 0cm
                0cm,width=8cm]{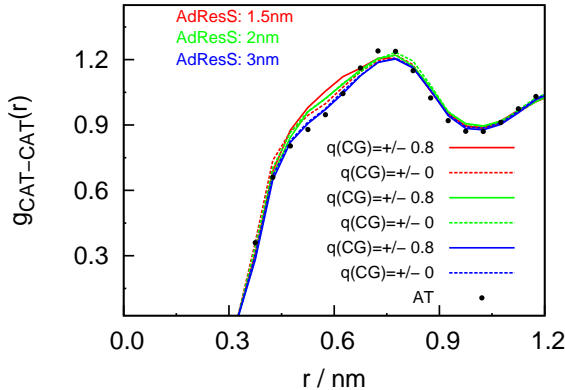}
                \caption{The variation of the cation-cation radial distribution
                  profile  calculated in the atomistic region of AdResS for both
                coarse-grained models and compared with the corresponding
                quantity calculated in the equivalent region of a full
              atomistic simulation.}
                \label{fig:fig12}
                \end{figure}

\begin{figure}[htbp]
	        \centering
                \includegraphics[clip=true,trim=0cm 0cm 0cm
                0cm,width=8cm]{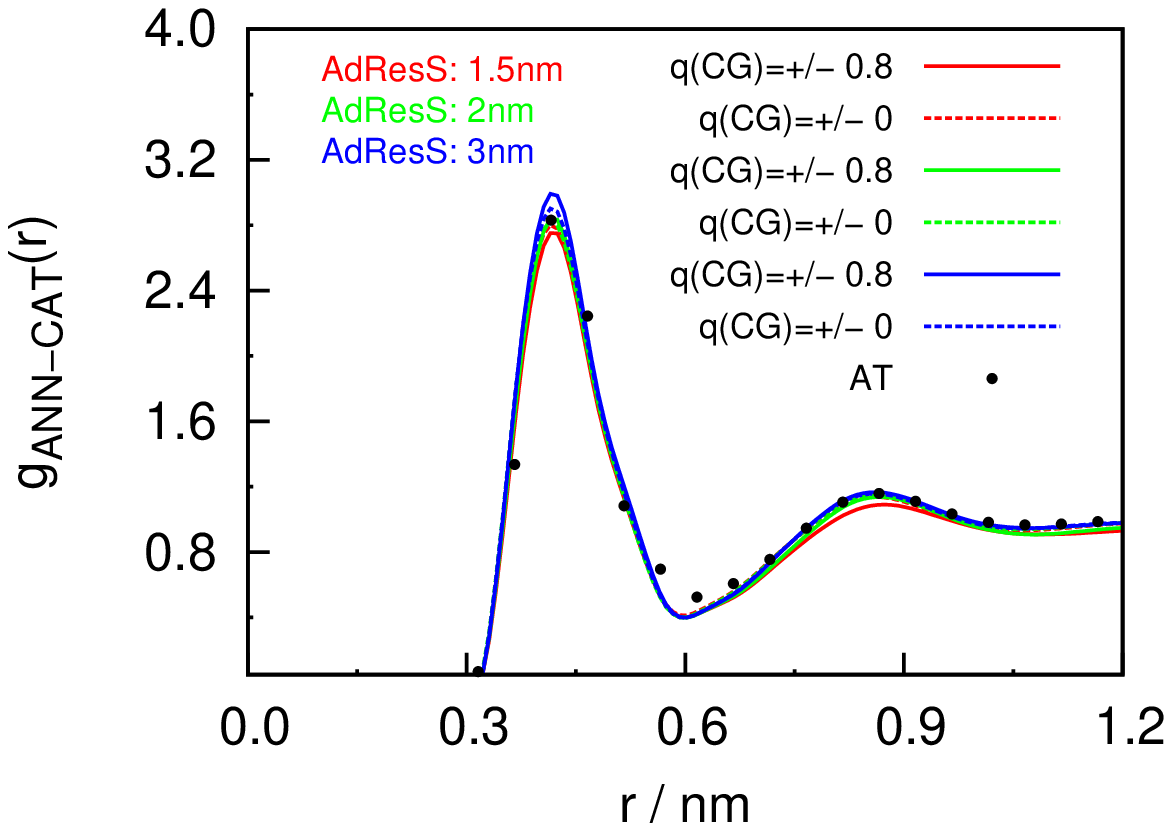}
                \caption{The anion-cation radial distribution
                  function calculated in the atomistic region of AdResS for both
                coarse-grained models and compared with the corresponding
                quantity calculated in the equivalent region of a full
              atomistic simulation.}
                \label{fig:fig13}
                \end{figure}

\begin{figure}[htbp]
	        \centering
                \includegraphics[clip=true,trim=0cm 0cm 0cm
                0cm,width=8cm]{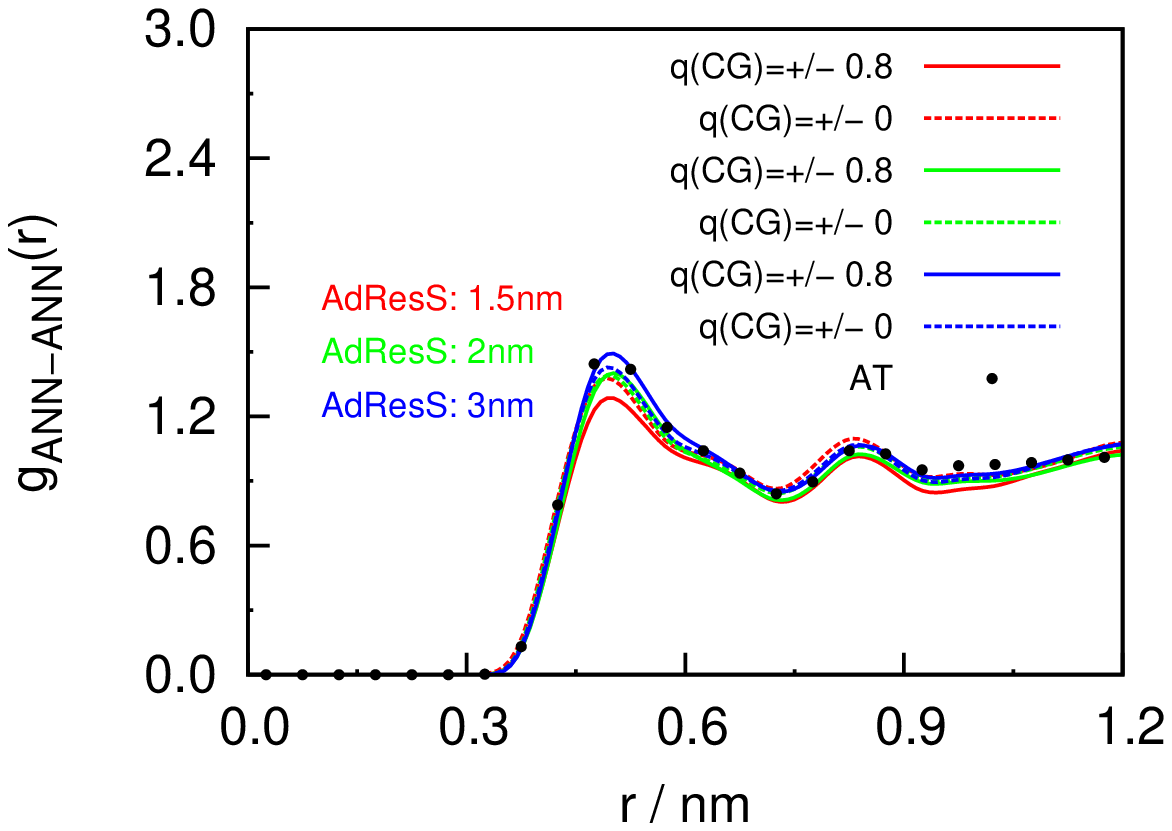}
                \caption{The anion-anion radial distribution
                  function calculated in the atomistic region of AdResS for both
                coarse-grained models and compared with the corresponding
                quantity calculated in the equivalent region of a full
              atomistic simulation.}
                \label{fig:fig14}
                \end{figure}

\begin{figure}[htbp]
	        \centering
                \includegraphics[clip=true,trim=0cm 0cm 0cm
                0cm,width=8cm]{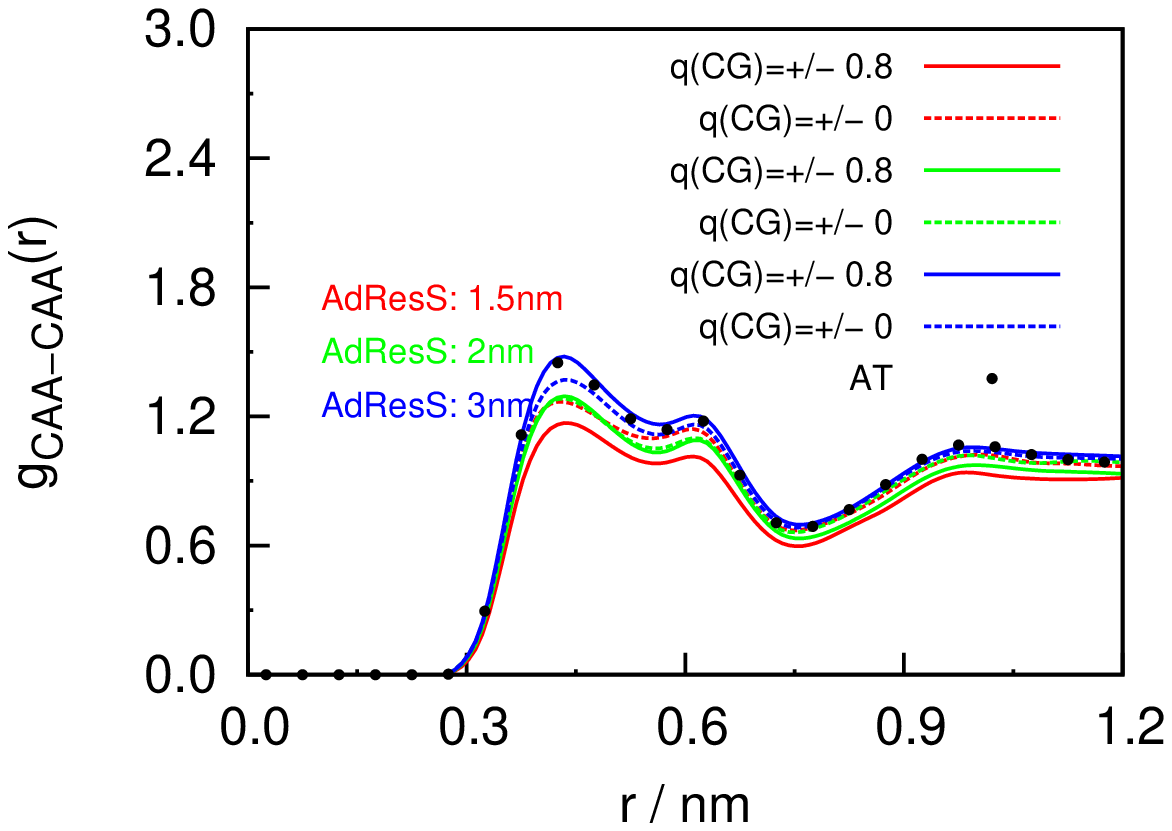}
                \caption{The carbon-carbon radial distribution
                  profile  calculated in the atomistic region of AdResS for both
                coarse-grained models and compared with the corresponding
                quantity calculated in the equivalent region of a full
              atomistic simulation.}
                \label{fig:fig15}
                \end{figure}

\begin{figure}[htbp]
	        \centering
                \includegraphics[clip=true,trim=0cm 0cm 0cm
                0cm,width=8cm]{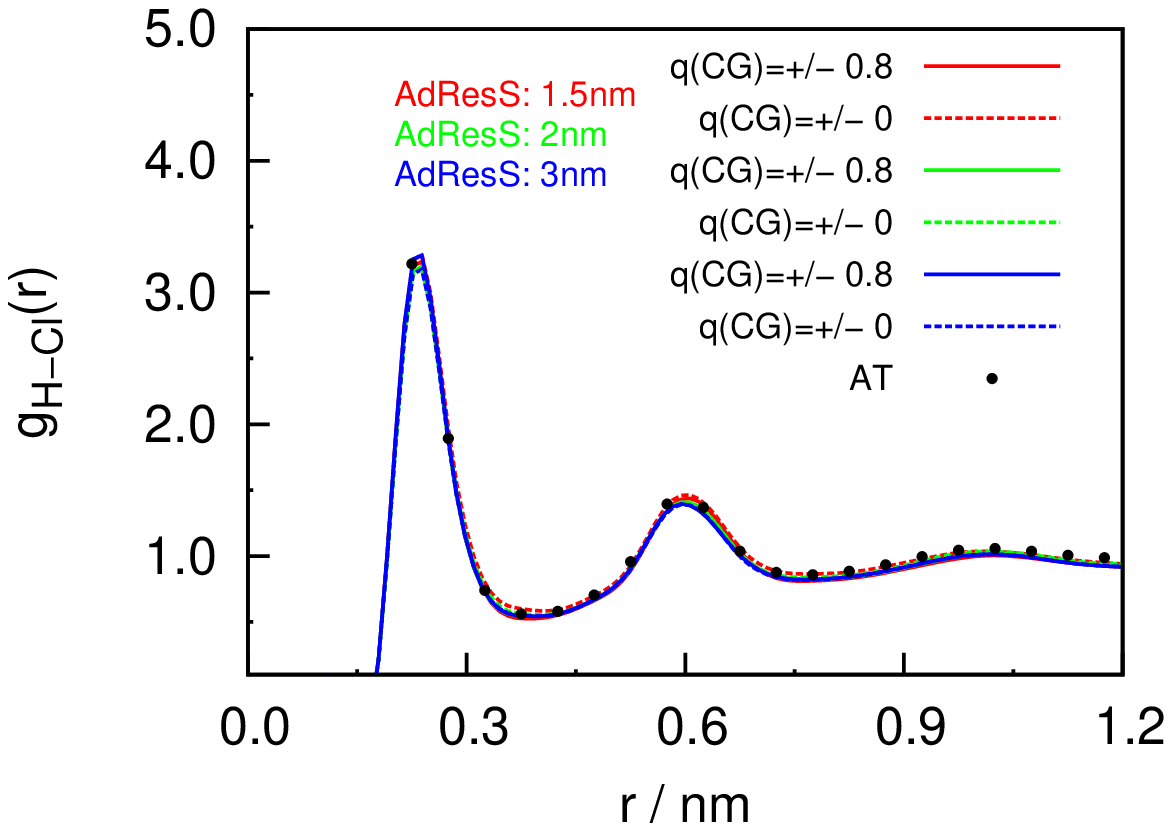}
                \caption{The hydrogen-chloride radial distribution
                  function  calculated in the atomistic region of AdResS for both coarse-grained models and compared with the corresponding
                quantity calculated in the equivalent region of a full
              atomistic simulation.}
                \label{fig:fig16}
                \end{figure}

\begin{figure}[htbp]
	        \centering
                \includegraphics[clip=true,trim=0cm 0cm 0cm
                0cm,width=8cm]{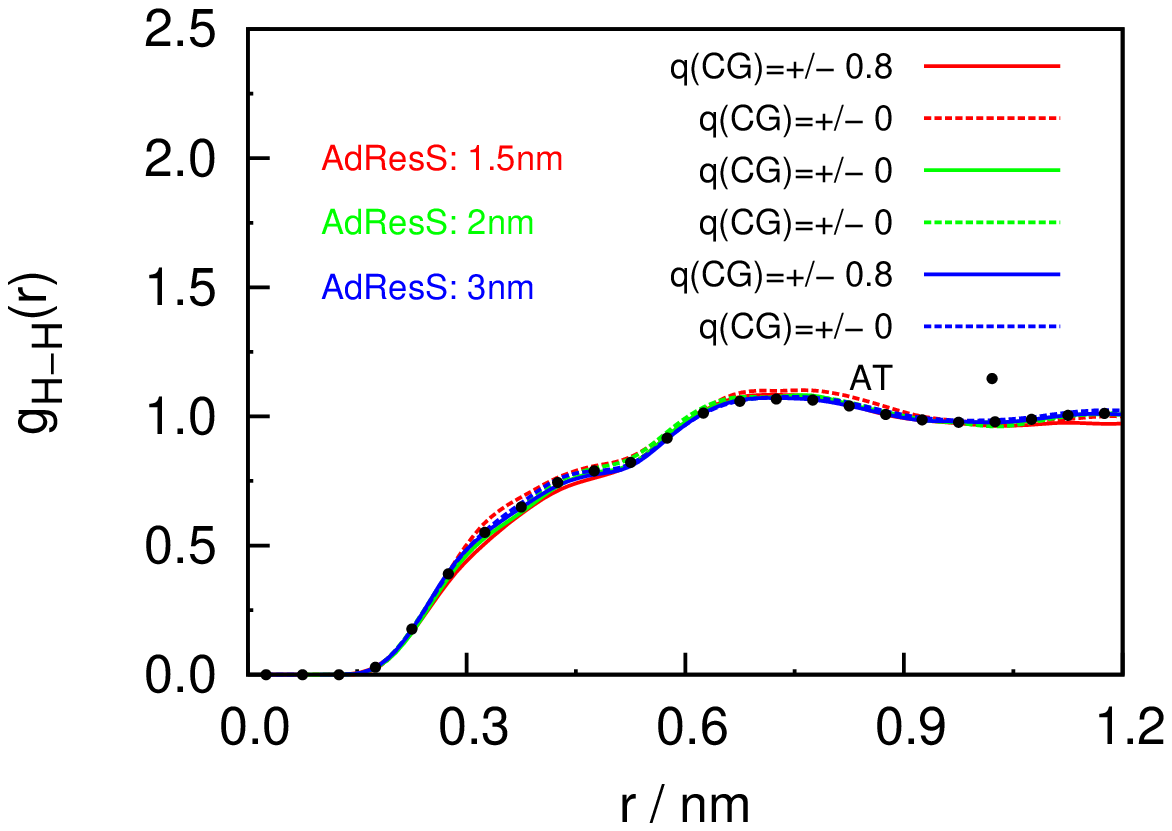}
                \caption{The hydrogen-hydrogen radial distribution
                  function  calculated in the atomistic region of AdResS for both coarse-grained models and compared with the corresponding
                quantity calculated in the equivalent region of a full
              atomistic simulation.}
                \label{fig:fig17}
                \end{figure}

Figures
\ref{fig:fig12}, \ref{fig:fig13}, \ref{fig:fig14}, \ref{fig:fig15},
\ref{fig:fig16}, \ref{fig:fig17}, show results of various radial distribution
functions. The different coarse-grained models, as already seen for the other
quantities in the previous section, do not play a key role. The {results 
show} that for an atomistic region of $3.0\, {\rm  nm}$ the agreement with the reference
full atomistic simulation is rather satisfactory and that we can safely take
$3.0 \, {\rm  nm}$ as the upper bound of locality{. However,} even for $1.5 \, {\rm  nm}$ 
the deviation from the data of reference is not dramatic{, so 
that, although with somewhat reduced accuracy, even an atomistic region of $1.5 \, {\rm nm}$ 
should be sufficient for the purpose.} 
This result {seems} interesting because $1.5 \, {\rm nm}$ corresponds to the size of 
few EMIM cations (Figure \ref{fig:fig10} shows that there are only 70 ion pairs) and thus 
the level of locality is considerable. 

{Note also that similar results are obtained with a coarse-grained model involving
uncharged particles. As a consequence, the idea of {\bf explicit charge-charge (atom-atom)} long range 
electrostatics playing a specific key role in the atomistic structure of the liquid seems 
refuted.} {In any case, it must be also added that even for a CG model with neutral sites, the long-range feature of the electrostatic interaction is effectively incorporated in the CG potentials, therefore, a good reproduction of structural properties by a CG model with neutral sites cannot defeat the general importance of the long-range feature of the electrostatic interaction in IL systems, but, as specified above, it can only point out that the corresponding (explicit) charge atomistic degrees of freedom are not relevant}. In the next section we discuss this point further. 
\subsection{Modeling Perspectives}
\begin{figure}[htbp]
\centering
\includegraphics[clip=true,trim=0cm 0cm 0cm 0cm,width=8cm]{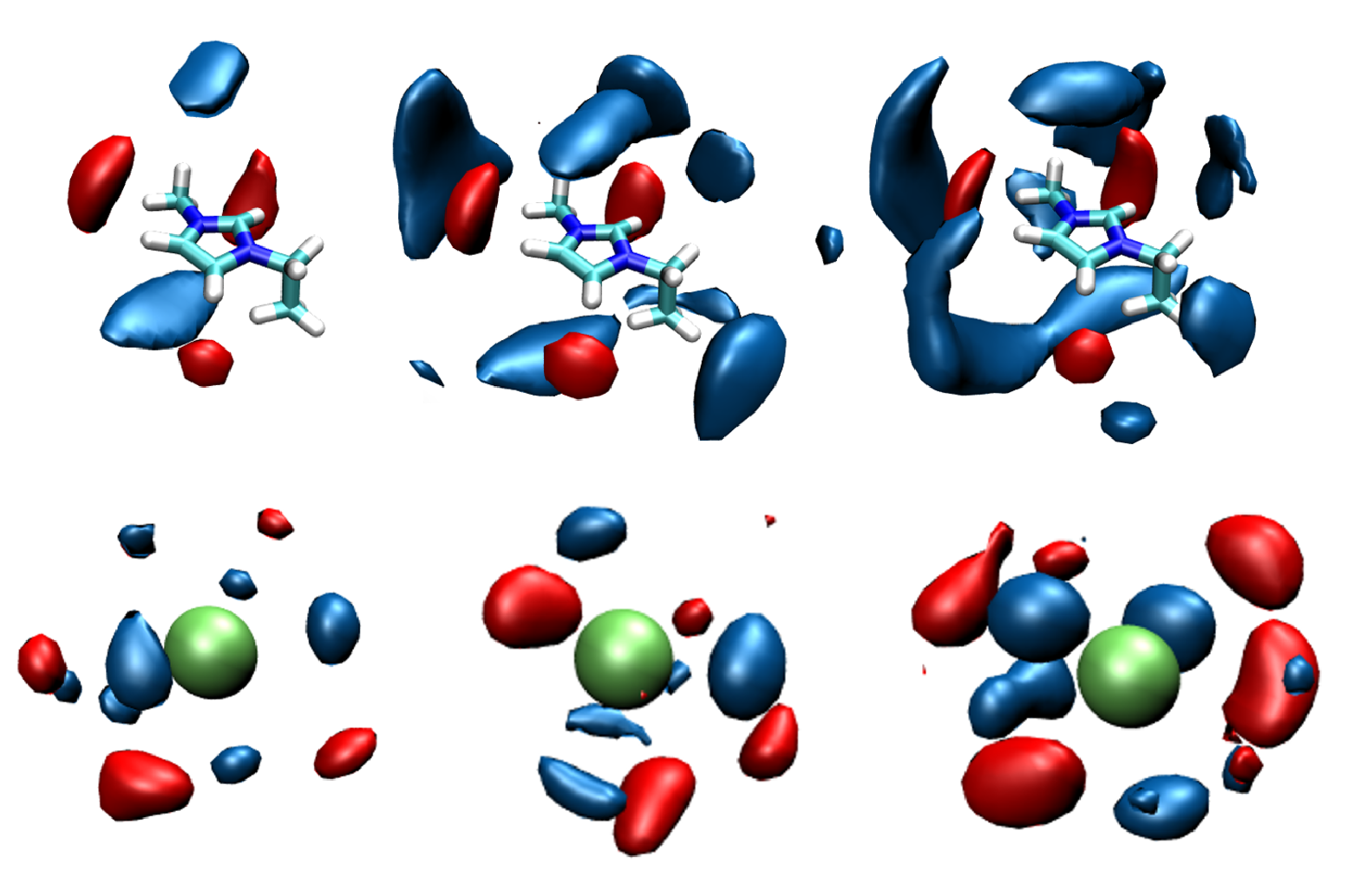}\\
\includegraphics[clip=true,trim=0cm 0cm 0cm 0cm,width=8cm]{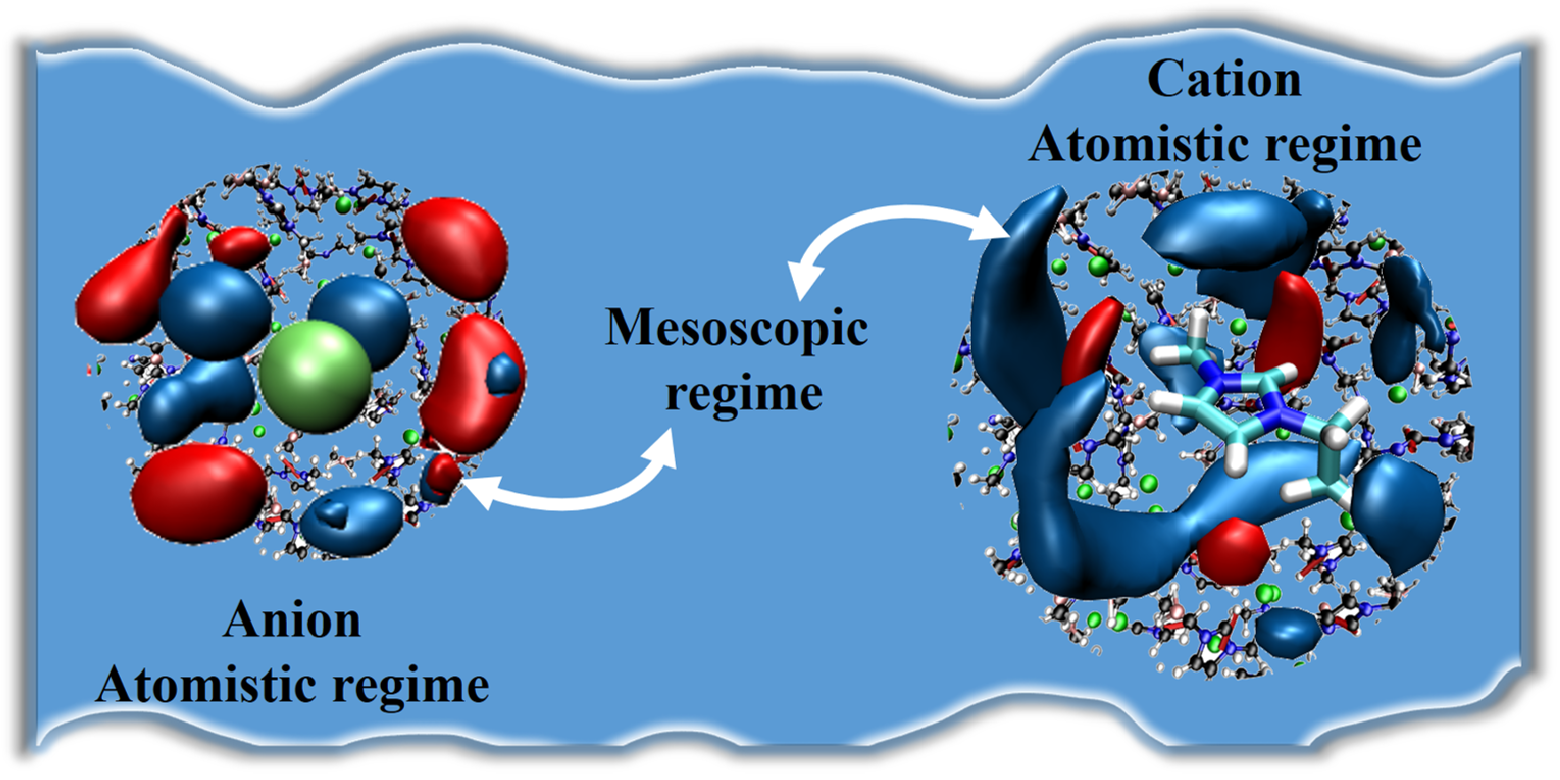}
\caption{Isosurfaces of the spatial probability density of anions (red) and cations (blue) 
as a function of the size of the atomistic regions (from $1.5\, {\rm nm}$ (left) to 
$3\, {\rm nm}$ (right)) in GC-AdResS simulations (with charge). The top and middle panels 
show the solvation environment of ions around the centered cations and anions, respectively. 
The bottom panel reports a pictorial representation of the essential solvation environment 
in space.}
\label{fig:fig18}
\end{figure}

The results shown in previous section suggest an overall scenario as that pictorially illustrated in Figure
\ref{fig:fig18}. One can reduce the model of an IL to the essential aspects by defining open atomistic islands centered around an anion
and a cation. These islands are embedded in a macroscopic environment characterized only by
macroscopic quantities, e.g. temperature, pressure, density, viscosity etc etc..
The representation of Figure \ref{fig:fig18} resembles closely the idea of ILs characterized 
by the rattling of ions in ion cages proposed in the literature
\cite{thar,maginn,delpopolo,turton,carrete}. 
Our method allows {us} to go even further and define the minimal size of the cages.
In general our findings provide justification to a strategy of study based on reducing the 
atomistic analysis to small regions (for atomistic properties) and only consider the link of 
such regions with the reservoir for properties at a larger scale. This kind of analysis would
be extremely useful for {designing ILs at the molecular level specified by clear
structure-function relations}. Our findings also strengthen the conclusions of previous theoretical 
work regarding the general local character in space of ionic liquids \cite{wv05,wzdbh11}. 
{It suggests} that by defining these spherical atomistically independent regions one can 
build physically well founded coarse-grained models of units larger than the molecular size and 
containing an entire group of anions and cations. {An example would be the} solvation 
environment for an anion and cation sketched in Figure\ref{fig:fig18}. {Such} an idea has 
already been successfully used for other systems {such} as liquid water \cite{matejlast}. 

\section{Conclusion}
We have tested the spatial locality of 1-ethyl 3-methyl imidazolium chloride liquid by employing the 
Grand Canonical Adaptive Resolution Molecular Dynamics technique. By calculating various radial 
distribution functions between anions and cations as a function of the size of the atomistic region 
and comparing them with the equivalent quantities in a full atomistic simulation we could conclude 
that even for a spherical region of radius $1.5\, {\rm  nm}$ the atomistic degrees of freedom of the 
bulk do not play a relevant role. Our study can be seen in different perspectives (a) as a further 
argument that ionic liquids {possess} an inherent locality in the spatial structure, {a}
hypothesis put forward in previous work \cite{wv05,wzdbh11}, {and} (b) as a modeling tool 
for {designing ILs at the molecular level} in the framework of structure-function relations.
{However, it must be underlined that our conclusions for other ionic liquids at this stage can only be qualitative and used a starting point for further studies. In fact the current systems is characterized by a relatively short alkyl side chain, thus the level of locality may change in a significant way for systems with longer alkyl chains. In this perspective, the next step will consist of repeating this analysis for a systematic 
comparison between different ILs aiming at understanding how a chemical modifications (e.g. the length of the of alkyl side chain in the cations) 
influence the larger scale behavior of the liquid and, in particular, its degree of locality. This 
study will be considered in future publications.}

Further interesting future perspectives may be offered by a recent proposal of GC-AdResS scheme for 
molecules with electrons, treated at {the} quantum level \cite{luigilast}{. Such an 
approach may pave the way for direct tests} of locality at the very accurate chemical level and thus 
the open the possibility of a precise chemical manipulation of the molecules and a direct testing of 
its {consequences within a larger scale environment}.

\section*{Acknowledgment}
This research has been funded by Deutsche Forschungsgemeinschaft (DFG) through
{grants CRC 1114 ``Scaling Cascades in Complex Systems'', project C01}.
This work has also received funding from the European Union's Horizon 2020 research
and innovation program under the grant agreement No. 676531 (project E-CAM). We would like to thank Barbara Kirchner for a critical reading of the manuscript and the members of her group for providing technical assistance regarding the choice and implementation of the force fields used.
\cleardoublepage
\def\bibsection{\section*{\refname}}
\end{document}